\documentclass[useAMS,usenatbib]{mn2e}

\usepackage{graphicx}
\usepackage{natbib}
\bibpunct{(}{)}{;}{a}{}{,}
\usepackage{txfonts}

\title[Polysulfanes on interstellar grains as a possible reservoir of interstellar sulphur]{Polysulfanes on interstellar grains as a possible reservoir of interstellar sulphur}
\author[C. Druard and V. Wakelam]{C. Druard$^{1,2}$ and V. Wakelam$^{1,2}$\\
$^{1}$Univ. Bordeaux, LAB, UMR 5804, F-33270, Floirac, France\\
$^{2}$CNRS, LAB, UMR 5804, F-33270, Floirac, France }
\begin{document}

\date{Accepted XXX Received XXX; in original form XXX}

\pagerange{\pageref{firstpage}--\pageref{lastpage}} \pubyear{2012}

\maketitle

\label{firstpage}

\begin{abstract}
The form of depleted sulphur in dense clouds is still unknown. Until now, only two molecules, OCS and SO$_2$, have been detected in interstellar ices but cannot account for the elemental abundance of sulphur observed in diffuse medium. Chemical models suggest that solid H$_2$S is the main form of sulphur in denser sources but observational constraints exist that infirm this hypothesis. We have used the Nautilus gas-grain code in which new chemical reactions have been added, based on recent experiments of H$_2$S ice irradiation with UV photons and high energy protons. In particular, we included the new species S$_n$, H$_2$S$_n$ and C$_2$S. We found that at the low temperature observed in dense clouds, i.e. 10~K, these new molecules are not efficiently produced and our modifications of the network do not change the previous predictions. At slightly higher temperature, 20~K in less dense clouds or in the proximity of protostars, H$_2$S abundance on the surfaces is strongly decreased in favor of the polysulfanes H$_2$S$_3$. Such a result can also be obtained if the diffusion barriers on the grains are less important. In the context of the life cycle of interstellar clouds and the mixing between diffuse and denser parts of the clouds, the depletion of sulphur in the form of polysulfanes or other sulphur polymers, may have occurred in regions where the temperature is slightly higher than the cold inner parts of the clouds.

\end{abstract}


\section{Introduction}

 Contrary to other elements \citep[see][]{2009ApJ...700.1299J}, the gas-phase abundance of atomic sulphur, in the diffuse medium, is observed to be constant with cloud density, around its cosmic value of $\sim 1.5\times 10^{-5}$ (compared to H) \citep{1994ApJ...430..650S}. In dense clouds (with densities above $10^4$~cm$^{-3}$), the sum of the abundances of molecules containing sulphur is a small fraction of the cosmic abundance of S \citep[about $10^{-3}$, see][]{1992IAUS..150..171O,2000ApJ...542..870D}. To reproduce such low S-bearing abundances, one needs to use a very small initial abundance of atomic sulphur of a few $10^{-8}$ \citep[see for example][]{2008ApJ...680..371W}. This suggests that sulphur depletion occurs very efficiently and rapidly between the diffuse and dense phases of the cloud life leading to an unknown, still unobserved, reservoir of sulphur on grains. When an atom of sulphur sticks to a grain, models predicts that it would be hydrogenated to form H$_2$S \citep{2007A&A...467.1103G}. H$_2$S has however never been detected in interstellar ices and its abundance has been estimated to be smaller than $5\times 10^{-8}$ (/H) \citep{1998ARA&A..36..317V,2011A&A...536A..91J}. Solid OCS has been observed with an abundance of about $5\times 10^{-8}$ \citep{1997ApJ...479..839P}. SO$_2$ may also be present with a similar abundance according to \citet{1997A&A...317..929B} and \citet{2009ApJ...694..459Z} . Solid H$_2$SO$_4$ has been proposed as a possible reservoir but no detection have confirmed this hypothesis \citep{2003MNRAS.341..657S}. Finally, iron sulfide FeS has been observed in protoplanetary disks by \citet{2002Natur.417..148K}. The depletion of sulphur in diffuse clouds does however not follow the one of iron so that the iron left in dense clouds cannot account for such a depletion of sulphur. 

Laboratory experiments have recently brought some new insights into the sulphur problem. \citet{garozzo} and \citet{2011A&A...536A..91J} have independently recently published experimental results of the irradiation of H$_2$S interstellar analogs ices by high energy protons and UV photons respectively. Both studies found that solid H$_2$S was easily destroyed in favor of other species such as OCS, SO$_2$, CS$_2$, H$_2$S$_2$ etc. Based on these results, we have revisited the sulphur depletion problem in dense clouds introducing new mechanisms for the formation of polysulfanes (H$_2$S$_n$), CS$_2$ and sulphur polymers (S$_n$) on the grains. The model, parameters and modifications of the chemical network are described in section~\ref{model}. The results of our model are presented in section~\ref{results}. We conclude in the last section. 

\section{Model description}\label{model}

We have used the chemical model Nautilus, which computes the abundances of chemical species as a function of time in the gas-phase and at the surface of interstellar grains. Starting from an initial composition, the model solves kinetic equations that describe how species abundances are modified by the various reactions destroying or forming them. These reactions are typical gas-phase reactions \citep[see][for a description of these processes]{2010SSRv..156...13W}, interactions between the gas and dust surfaces (adsorption of gas-phase species on grains and evaporation processes through various mechanisms) and surface reactions. Surface reactions stand for the reactions between two species on grain surfaces (this process implies that species move on the surface to meet), as well as dissociation of species on surfaces by external UV photons and UV photons produced by the interaction with cosmic-rays (see section~\ref{network}). Details on these processes and on the equations used to describe them can be found in \citet{2009A&A...493L..49H} and in \citet{2010A&A...522A..42S}.

\subsection{Chemical network}
The chemical network used for this work is the same as the one used by \citet{2011A&A...530A..61H}. It contains 6142 reactions, of which 4394 are pure gas-phase reactions and 1748 are grain-surface and gas-grain interaction reactions. The model follows the chemistry of 458 gas-phase species and 195 species on grains. The gas-phase network is based on the osu-08-2009 database (http://www.physics.ohio-state.edu/$\sim$eric/research.html) with updates according to the recommendations from the experts of the KIDA database (http://kida.obs.u-bordeaux1.fr). 

To this network, we added reactions for a new species: CS$_2$ (as well as CS$_2^+$, HCS$_2^+$ and CS$_2$ on the surface). Since CO$_2$ and CS$_2$ are isoelectronic, we have based the new reactions for CS$_2$ on those of CO$_2$. Rate constants for these reactions have been taken as mean values among the rates of the same type of reactions. We also added new reactions of formation on the grains for S-bearing molecules for OCS, SO$_2$ and H$_2$S$_2$. Formation of sulphur polymers S$_n$ (n = 2-8) and polysulfane H$_2$S$_3$ on the grains was also added, for these molecules could be good candidates to a sulphur refractory residue \citep{polysulfanes}. This extends the network to 666 species (461 for the gas-phase and 203 on the grains) and 6215 reactions. The reactions that we added to the network are listed in Tables~\ref{table1}, \ref{table2}, \ref{table3} and \ref{table4}, and the full network is available on the KIDA website (http://kida.obs.u-bordeaux1.fr/models).

\begin{table}[h]
\caption{Formation reactions for the main S-bearing species\label{table1}}
\centering 
\begin{tabular}{lllll}
	\multicolumn{5}{c}{Grain surface reactions} \\
	\hline\hline
	JHS & $+$ & JHS & $\longrightarrow$ & JH$_2$S$_2$\\
	JH$_2$S & $+$ & JS & $\longrightarrow$ & JH$_2$S$_2$\\
	JHS$_2$ & $+$ & JH & $\longrightarrow$ & JH$_2$S$_2$\\
	JH$_2$ & $+$ & JS$_2$ & $\longrightarrow$ & JH$_2$S$_2$\\
	JSO & $+$ & JC & $\longrightarrow$ & JOCS\\
	JS & $+$ & JO$_2$ & $\longrightarrow$ & JSO$_2$\\
\end{tabular}
\end{table}

\begin{table}[h]
\caption{S-polymers formation reactions on the grains  (where the larger polymer formed is S$_8$).\label{table2}}
\centering 
\begin{tabular}{lllll}
	\multicolumn{5}{c}{Grain surface reactions} \\
	\hline\hline
	JS$_n$ & $+$ & JS & $\longrightarrow$ & JS$_{n+1}$\\
	JS$_n$ & $+$ & JS$_n$ & $\longrightarrow$ & JS$_{2n}$\\
	\multicolumn{5}{c}{...} \\
\end{tabular}
\end{table}

\begin{table}[h]
\caption{Formation and destruction reactions of the polysulfanes on the grains (where the larger considered polysulfane is H$_2$S$_3$).\label{table3}}
\centering 
\begin{tabular}{lllllll}
	\multicolumn{7}{c}{Grain surface reactions} \\
	\hline\hline
	JH$_2$S$_2$ & $+$ & JS & $\longrightarrow$ & JH$_2$S$_3$ &  &   \\
	JHS$_2$ & $+$ & JHS & $\longrightarrow$ & JH$_2$S$_3$ &  &   \\
	\multicolumn{7}{c}{Grain surface CR photodissociation} \\
	\hline\hline
	JH$_2$S$_3$ &  &  & $\longrightarrow$ & JHS & $+$ & JHS$_2$ \\
	\multicolumn{7}{c}{Grain surface UV photodissociation} \\
	\hline\hline
	JH$_2$S$_3$ &  &  & $\longrightarrow$ & JHS & $+$ & JHS$_2$ \\
\end{tabular}
\end{table}

\begin{table}[h]
\caption{Gas-grain CS$_2$ chemistry\label{table4}}
\centering 
\begin{tabular}{lllllllll}
\multicolumn{9}{c}{Grain surface reactions} \\
\hline\hline
JS & $+$ & JCS & $\longrightarrow$ & JCS$_2$ &  & & & \\ 
JS & $+$ & JHCS & $\longrightarrow$ & JCS$_2$ & $+$ & JH & & \\ 
JHS & $+$ & JCS & $\longrightarrow$ & JCS$_2$ & $+$ & JH & & \\ 
\multicolumn{9}{c}{Thermal evaporation} \\
\hline\hline
JCS$_2$ &  &  & $\longrightarrow$ & CS$_2$ &  &  & & \\ 
\multicolumn{9}{c}{Cosmic ray desorption} \\
\hline\hline
JCS$_2$ &  &  & $\longrightarrow$ & CS$_2$ &  &  & & \\ 
\multicolumn{9}{c}{Grain surface CR photodissociation} \\
\hline\hline
JCS$_2$ &  &  & $\longrightarrow$ & JCS & $+$ & JS & &  \\ 
\multicolumn{9}{c}{Grain surface UV photodissociation} \\
\hline\hline
JCS$_2$ &  &  & $\longrightarrow$ & JCS & $+$ & JS & & \\
\multicolumn{9}{c}{Adsorption} \\
\hline\hline
CS$_2$ &  &  & $\longrightarrow$ & JCS$_2$ &  &  & & \\
\multicolumn{9}{c}{Gas phase CR photodissociation} \\
\hline\hline
CS$_2$ &  &  & $\longrightarrow$ & CS & $+$ & S &  &  \\
\multicolumn{9}{c}{Gas phase reactions} \\
\hline\hline
	C$^+$ & $+$ & CS$_2$ & $\longrightarrow$ & CS$^+$ & $+$ & CS &  & \\
	H$^+$ & $+$ & CS$_2$ & $\longrightarrow$ & HCS$^+$ & $+$ & S &  &  \\
	He$^+$ & $+$ & CS$_2$ & $\longrightarrow$ & C$^+$ & $+$ & S$_2$ & $+$ & He\\
	He$^+$ & $+$ & CS$_2$ & $\longrightarrow$ & S$^+$ & $+$ & CS & $+$ & He \\
	He$^+$ & $+$ & CS$_2$ & $\longrightarrow$ & CS$^+$ & $+$ & S & $+$ & He \\
	He$^+$ & $+$ & CS$_2$ & $\longrightarrow$ & S$_2^+$ & $+$ & C & $+$ & He \\
	He$^+$ & $+$ & CS$_2$ & $\longrightarrow$ & CS$_2^+$ & $+$ & He &  &  \\
	N$^+$ & $+$ & CS$_2$ & $\longrightarrow$ & CS$^+$ & $+$ & NS &  & \\
	N$^+$ & $+$ & CS$_2$ & $\longrightarrow$ & CS$_2^+$ & $+$ & N &  &  \\
	O$^+$ & $+$ & CS$_2$ & $\longrightarrow$ & SO$^+$ & $+$ & CS &  &  \\
	H$_2^+$ & $+$ & CS$_2$ & $\longrightarrow$ & CS$^+$ & $+$ & H$_2$S &  & \\
	H$_2^+$ & $+$ & CS$_2$ & $\longrightarrow$ & CS$_2^+$ & $+$ & H$_2$ &  &  \\
	H$_2^+$ & $+$ & CS$_2$ & $\longrightarrow$ & HCS$_2^+$ & $+$ & H &  &  \\
	OH$^+$ & $+$ & CS$_2$ & $\longrightarrow$ & HCS$_2^+$ & $+$ & O &  & \\
	H$_3^+$ & $+$ & CS$_2$ & $\longrightarrow$ & HCS$_2^+$ & $+$ & H2 &  & \\
	H & $+$ & CS$_2$ & $\longrightarrow$ & CS & $+$ & HS &  & \\
	S & $+$ & HCS & $\longrightarrow$ & H & $+$ & CS$_2$ &  & \\
	CS & $+$ & HS & $\longrightarrow$ & H & $+$ & CS$_2$ &  &  \\
	CS$_2^+$ & $+$ & e$^-$ & $\longrightarrow$ & S & $+$ & CS &  & \\
	CS$_2^+$ & $+$ & H & $\longrightarrow$ & H$^+$ & $+$ & CS$_2$ &  &  \\
	CS$_2^+$ & $+$ & H & $\longrightarrow$ & HCS$^+$ & $+$ & S &  & \\
	CS$_2^+$ & $+$ & O & $\longrightarrow$ & O$^+$ & $+$ & CS$_2$ &  & \\
	CS$_2^+$ & $+$ & O & $\longrightarrow$ & SO$^+$ & $+$ & CS &  & \\
	CS$_2^+$ & $+$ & H$_2$ & $\longrightarrow$ & HCS$_2^+$ & $+$ & H &  &  \\
	CS$_2^+$ & $+$ & O$_2$ & $\longrightarrow$ & O$_2^+$ & $+$ & CS$_2$ &  & \\
	CS$_2^+$ & $+$ & H$_2$O & $\longrightarrow$ & H$_2$O$^+$ & $+$ & CS$_2$ &  &  \\
	CS$_2^+$ & $+$ & H$_2$O & $\longrightarrow$ & HCS$_2^+$ & $+$ & OH &  &  \\
	CS$_2^+$ & $+$ & H$_2$S & $\longrightarrow$ & H$_2$S$^+$ & $+$ & CS$_2$ &  &  \\
	HCS$_2^+$ & $+$ & C & $\longrightarrow$ & CH$^+$ & $+$ & CS$_2$ &  &  \\
	HCS$_2^+$ & $+$ & O & $\longrightarrow$ & HCS$^+$ & $+$ & SO &  &  \\
	HCS$_2^+$ & $+$ & CO & $\longrightarrow$ & HCS$^+$ & $+$ & OCS &  &  \\
	HCS$_2^+$ & $+$ & H$_2$O & $\longrightarrow$ & H$_3$O$^+$ & $+$ & CS$_2$ &  &  \\
	HCS$_2^+$ & $+$ & C$_2$H$_2$ & $\longrightarrow$ & C$_2$H$_3$+ & $+$ & CS$_2$ &  & \\
	HCS$_2^+$ & $+$ & CH$_4$ & $\longrightarrow$ & CH$_5$+ & $+$ & CS$_2$ &  &  \\
	HCS$_2^+$ & $+$ & e$^-$ & $\longrightarrow$ & CS & $+$ & H & $+$ & S \\
	HCS$_2^+$ & $+$ & e$^-$ & $\longrightarrow$ & HS & $+$ & CS &  &  \\
	HCS$_2^+$ & $+$ & e$^-$ & $\longrightarrow$ & CS$_2$ & $+$ & H &  &  \\	
\multicolumn{9}{c}{Gas phase UV photodissociation} \\
\hline\hline
	CS$_2$ &  &  & $\longrightarrow$ & CS & $+$ & S &  & \\
\end{tabular}
\end{table}

\subsection{Model parameters} 
\label{cloud_cond}

For the simulations of the chemistry in dense clouds, the gas and dust temperatures are set to 10~K. The H density is $2 \times 10^{4}$~cm$^{-3}$, the visual extinction is 30 and the total H$_2$ cosmic ray ionization rate is $1.3\times 10^{-17}$ s$^{-1}$. All elements are initially in the atomic form and in the gas-phase (assuming that dense clouds are formed from diffuse medium), except for hydrogen, which is supposed to be entirely in the form of H$_2$ in the gas. Initial abundances are listed in Table~\ref{Abini}. Elemental abundance of sulphur is at its cosmic abundance of $1.5\times 10^{-5}$ \citep{1994ApJ...430..650S}. 

   \begin{table}
     \caption{Initial abundances \citep[see][for discussion on these values]{2011A&A...530A..61H}}
        \label{Abini} 
\centering
         \begin{tabular}{ll}
            \hline
            Element & Abundance (/H) \\
            \hline \hline
            H$_2$ & $0.5$\\
            He & $9 \times 10^{-2}$ $^{\mathrm{a}}$\\
            N & $6.2 \times 10^{-5}$ $^{\mathrm{b}}$\\
            O & $1.4 \times 10^{-4}$ $^{\mathrm{c}}$\\
            C$^+$ & $1.7 \times 10^{-4}$ $^{\mathrm{b}}$\\
            S$^+$ & $1.5 \times 10^{-5}$ $^{\mathrm{d}}$\\
            Si$^+$ & $8 \times 10^{-9}$ $^{\mathrm{e}}$\\
            Fe$^+$ & $3 \times 10^{-9}$ $^{\mathrm{e}}$\\
            Na$^+$ & $2 \times 10^{-9}$ $^{\mathrm{e}}$\\
            Mg$^+$ & $7 \times 10^{-9}$ $^{\mathrm{e}}$\\
            P$^+$ & $2 \times 10^{-10}$ $^{\mathrm{e}}$\\
            Cl$^+$ & $1 \times 10^{-9}$ $^{\mathrm{e}}$\\
            \hline
         \end{tabular}
         \begin{list}{}{}
\item[$^{\mathrm{a}}$] \citet{2008ApJ...680..371W}
\item[$^{\mathrm{b}}$] \citet{2009ApJ...700.1299J}
\item[$^{\mathrm{c}}$] \citet{2011A&A...530A..61H}
\item[$^{\mathrm{d}}$] See text
\item[$^{\mathrm{e}}$] Low metal elemental abundances, \citep[see][]{1982ApJS...48..321G}
\end{list}
            \end{table}
            
\subsection{Model parameters for surface reactions}
\label{network}

For the formation of molecules on the surfaces, there are two key processes for which we give some details: the dissociation of species at the surface of the grains by photons induced by cosmic-ray particles and the reactions occurring at the surfaces. In our model, the efficiency of these processes depends on a number of parameters that we describe now. The direct interaction of cosmic-ray particles with chemical species produces mostly ionization of H$_2$ and atoms \citep{1978ApJ...219..750C,1980ApJS...43....1P}. Ionization of H$_2$ by cosmic-rays produces high energy electrons that excites H$_2$. By de-excitation, H$_2$ then produces a UV radiation field inside dense clouds \citep{1983ApJ...267..603P,1989ApJ...347..289G}. The dissociation of molecules by these photons are usually called cosmic-rays induced photodissociations. Such processes are also supposed to be efficient on the molecules contained in grain ices and their efficiencies are taken to be the same as for gas-phase molecules as in \citet{2000MNRAS.319..837R}. Cations produced by photo-ionizations are assumed to recombine on the surfaces of the grains, which are mainly negatively charged. Rate coefficients for these reactions are computed by the formula: k$_{CRP} = \alpha_{CRP} \zeta$ (in s$^{-1}$). $\zeta$ is the total H$_2$ cosmic-ray ionization rate and $\alpha_{CRP}$ specific for each species \citep[see][]{1989ApJ...347..289G}. Table~\ref{CRrates} lists the $\alpha_{CRP}$ used in our model for the main S-bearing molecules, which are from the OSU database\footnote{http://www.physics.ohio-state.edu/$\sim$eric/research.html}. For H$_2$S$_3$ on the surfaces, we have used the same photodissociation rate as for H$_2$S$_2$. The products of the photodissociation of H$_2$S$_2$ on the surfaces have been changed compared to the gas-phase based on the new experiments by \citet{2011A&A...536A..91J}.

Reactions on grain surfaces between two chemical species are computed using equation 5 from \citet{1992ApJS...82..167H}:
\begin{equation}
R_{i,j} = \kappa_{i,j}(R_{diff,i}+R_{diff,j})N_i N_j n_d
\end{equation}
with $R_{i,j}$ the rate (in cm$^{-3}$ s$^{-1}$) of reaction between species $i$ and $j$ at the surface of the grains, $\kappa_{i,j}$ is the probability for the reaction to occur (one for exothermic reactions), $N_i$ and $N_j$ the number of species $i$ and $j$ at the surface, $n_d$ is the number density of grains (in cm$^{-3}$). $R_{diff}$ are the diffusion rates of the species at the surface of the grains \citep[see equations 2 to 4 from ][]{1992ApJS...82..167H}. This diffusion rate is proportional to $\exp(-E_b / kT_d)$ with $T_d$ the dust temperature and $E_b$ the potential energy barrier required to move from one site to the other on the surface of the grains \citep{1976RvMP...48..513W}. This last parameter is uncertain and is usually taken to be a fraction of the binding energies ($E_D$) of the species to the surface. Various values for this fraction have been used over the years, typically ranging from 0.3 \citep{1976RvMP...48..513W,1987ppic.proc..333T,1992ApJS...82..167H} to 0.77 \citep{2000MNRAS.319..837R}. Here we use a mean value of 0.5 as in \citet{2006A&A...457..927G}.

 \begin{table}
     \caption{$\alpha_{CRP}$ for the main S-bearing species on the grains.}
        \label{CRrates} 
\centering
         \begin{tabular}{lllllc}
            \hline
            \multicolumn{5}{c}{CR induced photodissociation reaction} & $\alpha_{CRP}$ \\
            \hline \hline
            JH$_2$S & $\longrightarrow$ & JH$_2$ & $+$ & JS & $5.15 \times 10^{3}$\\
			JH$_2$S & $\longrightarrow$ & JS & $+$ & 2JH & $8.50 \times 10^{2}$\\
			JH$_2$S & $\longrightarrow$ & JHS & $+$ & JH & $8.50 \times 10^{2}$\\            
            JH$_2$S$_2$ & $\longrightarrow$ & JS2 & $+$ & JH$_2$ & $1.50 \times 10^{3}$\\
             JH$_2$S$_3$ & $\longrightarrow$ & JHS$_2$ & $+$ & JHS & $1.50 \times 10^{3}$\\           
            JHS & $\longrightarrow$ & JH & $+$ & JS & $5.00 \times 10^{2}$\\
            JOCS & $\longrightarrow$  & JCO & $+$ & JS & $6.31 \times 10^{3}$\\
%            JOCS & $\longrightarrow$ 18 & JCO & $+$ & JS & $9.60 \times 10^{2}$\\
            JOCS & $\longrightarrow$ & JCS & $+$ & JO & $4.80 \times 10^{2}$\\
            JSO$_2$ & $\longrightarrow$ & JSO & $+$ & JO & $1.88 \times 10^{3}$\\
            \hline
         \end{tabular}
\end{table}

\section{Modelling results}
\label{results}

\subsection{Under dense cloud conditions}

\begin{figure}
	\resizebox{\hsize}{!}{\includegraphics{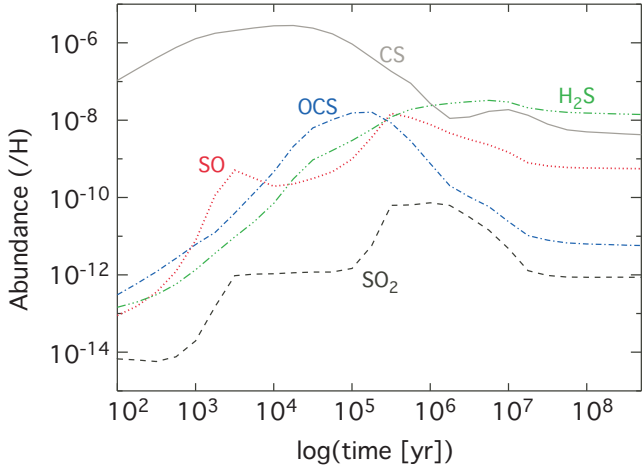}}
	\caption{Gas-phase abundances of the S-bearing species relative to H as a function of time for a gas and dust temperature of 10K.}
	\label{gaz_10K}
\end{figure}

\begin{figure}
	\resizebox{\hsize}{!}{\includegraphics{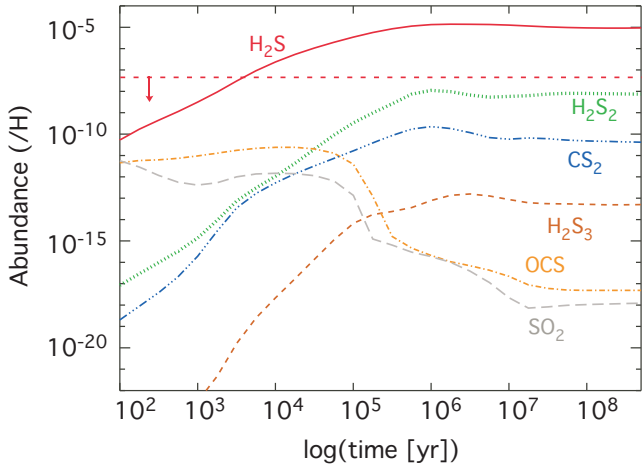}}
	\caption{Grain surface abundances of the S-bearing species relative to H as a function of time for a gas and dust temperature of 10~K. Red horizontal dashed line represents the upper limit on H$_2$S abundance observed in grain ices \citep{1998ARA&A..36..317V}.}
	\label{gaz_grain_10K}
\end{figure}

The model was used with the parameters described in the previous section. The abundances (compared to total H density) for the main S-bearing molecules in the gas-phase and on the grains are shown in Figs.~\ref{gaz_10K} and \ref{gaz_grain_10K} as a function of time. In the gas-phase, the predicted abundances of CS, SO and H$_2$S are much larger than the ones observed in the dense cloud L134N (North Peak). A comparison between the predicted (at $10^6$~yr) and observed abundances is listed in Table~\ref{AbL134N}. Observed abundances from several authors are given to see the level of uncertainties in the observations. In parallel, sulphur, which is initially in the atomic form in the gas-phase, depletes efficiently on grains and is hydrogenated to form H$_2$S. Hydrogen sulfide on grains is then the main reservoir of sulphur and its abundance predicted by the model is several orders of magnitude larger than the observational upper limit, similarly to previous modelings \citep[see for instance][]{2007A&A...467.1103G}. The other S-bearing molecules efficiently produced on the grains are H$_2$S$_2$ and CS$_2$ but with abundances of $1.1\times 10^{-8}$ and $2.2\times 10^{-10}$ respectively at $10^6$~yr. At low temperatures as 10~K, the only efficient grain surface reactions are hydrogenations, which reform H$_2$S when dissociated. The modifications that we have done on the chemical network do not affect the results here, except that OCS is less produced on the surface in favor of the newly introduced species CS$_2$. For other S-bearing molecules to form on the grains, the diffusivity of the species on the grains have to be increased.

%To improve the agreement between models and observations, smaller elemental abundances of S are usually used \citep{1982ApJS...48..321G}. With a sulphur elemental abundance of $8\times 10^{-8}$, we obtain the following gas-phase abundances at $10^6$~yr: $1.8\times 10^{-9}$ for CS, $8.3\times 10^{-10}$ for SO and $5.9\times 10^{-9}$ for H$_2$S. The CS and SO abundances are then closer to the observed ones but the gas-phase abundance of H$_2$S predicted by the model, which comes from the evaporation of H$_2$S formed on grains \citep[see also][]{1989MNRAS.237.1057H,2007A&A...467.1103G}, is still higher by a factor of 4 to 10 than the observed one. All these results suggest that H$_2$S ice cannot be the reservoir of sulphur. 

%The predicted abundances of S-bearing molecules in the gas-phase using this model (high elemental abundance of S) is shown in the appendix~\ref{append_B}.

\begin{table*}
	\caption{Abundances relative to H observed in L134N (North Peak).}
	\label{AbL134N} 
	\centering
		\begin{tabular}{llllll}
			\hline
			Element & \multicolumn{5}{c}{Abundances (/H) of the main gas-phase S-bearing species.} \\
            \hline \hline
             & Model & \citet{1989ApJ...345..828S} & \citet{1992IAUS..150..171O} & \citet{1989ApJ...345L..63M} & \citet{2000ApJ...542..870D} \\
            \hline
            CS & $4\times 10^{-8}$ & $3.4\times 10^{-10}$ & $5\times 10^{-10}$ & - & $8.3\times 10^{-10}$ \\
            SO & $1\times 10^{-8}$& $0.8-2.1 \times 10^{-9}$$^a$ & $1\times 10^{-8}$ & - & $1.6\times 10^{-9}$ \\
            SO$_2$ & $1\times 10^{-10}$ & $\leq1.4\times 10^{-10}$ & $2\times 10^{-9}$ & - & $\leq8\times 10^{-10}$ \\            
            OCS & $1\times 10^{-9}$ & - & $1\times 10^{-9}$ & - & - \\
            H$_2$S & $4\times 10^{-8}$ & - & $4\times 10^{-10}$ & $1.5\times 10^{-9}$ & - \\
            \hline
         \end{tabular}
         $^a$ from two different molecular lines.
\end{table*}
 
\subsection{Sensitivity to the model parameters} 

We now test sensitivity of the model results to some of the key parameters.

\subsubsection{The gas and dust temperature}

\begin{figure}
	\resizebox{\hsize}{!}{\includegraphics{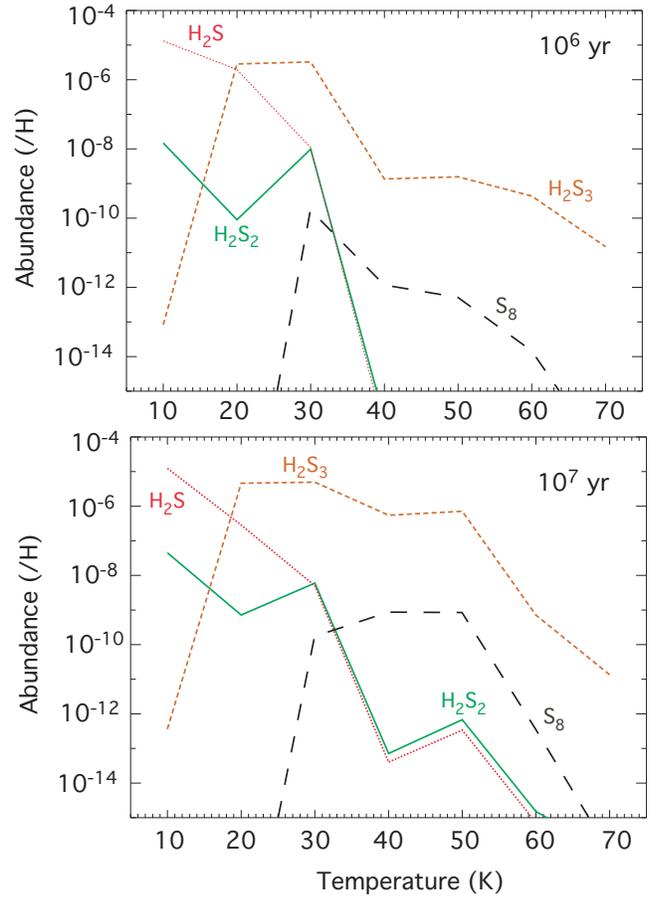}}
	\caption{Grain surface abundances of the H$_2$S, H$_2$S$_2$, H$_2$S$_3$ and S$_8$ as a function of temperature at two different times ($10^6$~yr on the top and $10^7$~yr at the bottom).}
	\label{ab_temp}
\end{figure}

\begin{figure}
	\resizebox{\hsize}{!}{\includegraphics{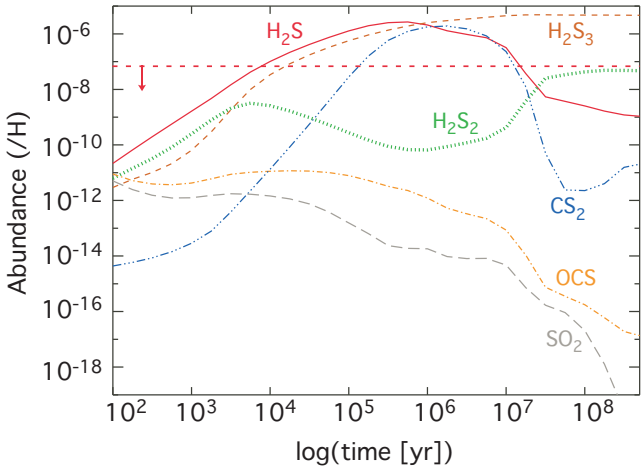}}
	\caption{Grain surface abundances of the S-bearing species relative to H as a function of time for a 20K gas-grain model. Red dashed horizontal line represents the upper limit on H$_2$S observed in grain ices \citep{1998ARA&A..36..317V}. }
	\label{gaz_grain_20K}
\end{figure}
 
Figure~\ref{ab_temp} shows the grain surface abundances of the H$_2$S, H$_2$S$_2$, H$_2$S$_3$ and S$_8$ at two different times ($10^6$ and $10^7$~yr) for temperatures  (gas and dust) from 10 to 70~K. At 20~K, half of the sulphur goes into H$_2$S$_3$ on the surface after $10^6$~yr and the abundance of H$_2$S is below $10^{-7}$ after $10^7$~yr (see also Fig.~\ref{gaz_grain_20K}). In this model, CS$_2$ is produced efficiently and its abundance is similar to that of H$_2$S for times between $10^6$ and $10^7$~yr. At a temperature of 30~K, the abundance of H$_2$S on the grains is at maximum $10^{-8}$ (at $10^6$~yr). Its gas-phase abundance is always smaller than $10^{-10}$. Above 30~K, the abundance of S-bearing molecules on the grains decreases because the evaporation temperature of atomic sulphur is reached. At these low densities ($10^4$~cm$^{-3}$), the main form of sulphur is then neutral atomic sulphur in the gas-phase. The molecule S$_8$ is efficiently produced by temperatures between 30 and 50~K.

 %
%Such temperatures are larger than what is observed in "typical" dense clouds such as L134N and are close to the dust temperature in massive star forming regions such as Orion or in less dense medium such as the border of molecular clouds.

\subsubsection{Energy barrier between two grain adjacent sites}

\begin{figure}
	\resizebox{\hsize}{!}{\includegraphics{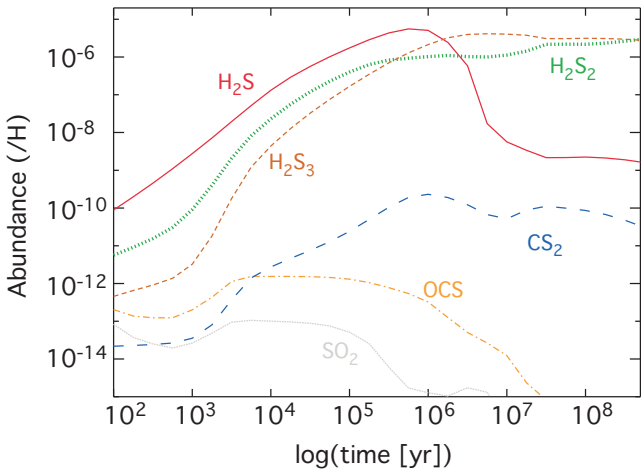}}
	\caption{Grain surface abundances of the S-bearing species relative to H as a function of time for $E_b = 0.3E_D$ (other parameters are the same as for Fig. \ref{gaz_grain_10K}). }
	\label{Eb_diff}
\end{figure}

Another parameter that influences directly the diffusion of the species on grains is the energy barrier between two grain adjacent sites $E_b$. As mentioned in the section 2.4, the value of this energy barrier is far from being well known. To test its importance, we have run the model described in section 2.3 (dust and gas temperature are 10~K) with an $E_b$ of 0.3$E_D$ (see section 2.4). The model results are shown in Fig.~ \ref{Eb_diff}. The destruction of H$_2$S on the surfaces is very efficient in that case in favor of the production of CS$_2$ and H$_2$S$_3$. After a few $10^6$~yr, H$_2$S$_3$ and H$_2$S$_2$ are the main S-bearing species on grains and the prediction abundance of H$_2$S is below its observational limit. In that case, CS$_2$ is not produced efficiently. In all the models shown in this section, the formation of pure sulphur polymers S$_n$ is very low and the predicted abundances are below the minimum values shown on the figures. 

\subsubsection{Other parameters}

\begin{figure}
	\resizebox{\hsize}{!}{\includegraphics{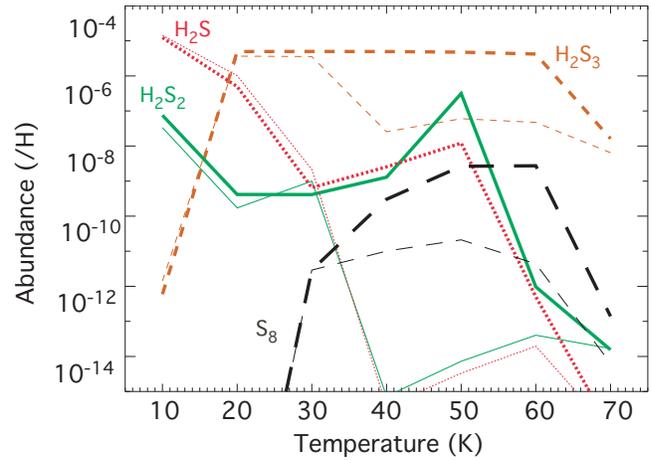}}
	\caption{Grain surface abundances of the H$_2$S, H$_2$S$_2$, H$_2$S$_3$ and S$_8$ as a function of temperature at two different times ($10^6$~yr thin line and $10^7$~yr thick line) for a H density of $2\times 10^5$~cm$^{-3}$.}
	\label{ab_temp_2e5}
\end{figure}

At a larger density of $2\times 10^5$~cm$^{-3}$, the results at 10~K are very similar to the lower density one. Between 20 and 60~K, solid H$_2$S$_3$ remains with a very high abundance similar to the neutral atomic sulphur in the gas. The molecule S$_8$ is efficiently produced for temperatures between 30 and 60~K. At 70~K, gas-phase S is the reservoir of sulphur. A figure similar to Fig.~\ref{ab_temp} for a density of $2\times 10^5$~cm$^{-3}$ is given in Fig.~\ref{ab_temp_2e5}. \\
The results presented here are not strongly sensitive to the cosmic-ray ionization rate. A ten times larger $\zeta$, with a temperature of 20~K,  accelerates the destruction of H$_2$S on the surface and favors the production of H$_2$S$_2$. H$_2$S$_3$ remains however the reservoir of sulphur after $5\times 10^5$~yr. A ten times smaller $\zeta$ only produces a smaller abundance of H$_2$S$_2$, below $10^{-10}$ for times between $10^6$ and $10^7$~yr. \\

\section{Conclusions}

%The depletion of sulphur in an unknown reservoir is a mandatory hypothesis to reproduce the abundances of S-bearing molecules observed in the gas-phase in dense clouds. 
Chemical models that take into account the formation of molecules on interstellar grains predict that sulphur would stick on the grains at high density and would be hydrogenated to form H$_2$S. The non detection of this species in interstellar ices tends to indicate that solid H$_2$S cannot be the reservoir of sulphur. The recent experiments from \citet{garozzo} and \citet{2011A&A...536A..91J} showed that H$_2$S is easily dissociated by high energy particles and UV photons, and that other molecules such as CS$_2$, polysulfanes (H$_2$S$_n$) and S$_n$ would be formed on the surfaces. Using a gas-grain model, in which new reactions have been introduced for CS$_2$, H$_2$S$_n$ and S$_n$, we show that we are able to produce large quantities of solid H$_2$S$_3$ and very low abundance of solid H$_2$S but only with temperatures higher than 10~K, which are probably more representative of star forming regions. The key parameter here is the diffusivity of the radical HS on the grains, which is limited at very low temperature. Significant amount of S$_8$ is produced at temperatures between 30 and 50-60~K for densities between $2\times 10^4$ and $2\times 10^5$~cm$^{-3}$. The results of our model, as for any chemical model, depends on the chemical reactions considered. More experience on the diffusion of S-bearing species on surfaces at low temperature and the formation of these chains would be a precious ally.   

The possibility to observe products of the destruction of solid H$_2$S$_n$ or S$_n$ in diffuse or warm medium could bring more stones to this problem. H$_2$S$_2$ is a symmetric molecule and the predicted lines in the millimeter range are weak \citep{1990JMoSp.141..265B}. Search for these lines, as well as S$_3$ and S$_4$ \citep{2005JChPh.123e4326T,2005ApJ...619..939G}, in the spectral survey of the low mass protostar IRAS16293-2422 \citep{2011A&A...532A..23C}, was unsuccessful (E. Caux private communication). The results of this modeling can be put in the context of the larger picture of interstellar dust cycle \citep{1998ApJ...499..267T} and the fact that the depletion of sulphur into H$_2$S$_n$ may have happened at an early phase of the cloud history when the dust temperature was larger than what is observed at the centre of these cold cores. It is unlikely that simple diffusion mechanisms at the surface of the grains could form pure sulphur polymers (S$_n$). The observation of OCS and SO$_2$ in interstellar ices may indicate however that other mechanisms could take place or that surface reactions are much more efficient than current models assume.

\section*{Acknowledgments}

Franck Hersant is thanked for his comments on this paper. VW is grateful to the French program PCMI for partial funding of this work. CD was funded by a 4 months master grant from the Laboratoire d'Astrophysique de Bordeaux and the University of Bordeaux 1. Some kinetic data we used have been downloaded from the online database KIDA (KInetic Database for Astrochemistry, http://kida.obs.u-bordeaux1.fr). The authors are acknowledge the anonymous referee for his suggestions, which improved the quality of the paper.

\bibliographystyle{mn2e}
\nocite{*}
\bibliography{biblio}

%\appendix
%\section{Reactions added to the network}

\end{document}